\def\kms{\hbox{km s$^{-1}$}}

\def\cm2{cm$^{-2}$}
\def\cm3{cm$^{-3}$}

\def\radec{\hbox{RA, Dec.(J2000)}}
\def\gra{$^{\circ}$}

\documentclass[fleqn,usenatbib]{mnras}

\usepackage[T1]{fontenc}
\usepackage{ae,aecompl}
\usepackage{soul}

\usepackage{graphicx}	
\usepackage{amsmath}	
\usepackage{amssymb}	

\title[Short title, max. 45 characters]{ Cyanoacetylene in the outflow/hot molecular  core G331.512-0.103}

\author[N. U. Duronea et al.]{
N. U. Duronea,$^{1}$\thanks{E-mail: duronea@iar.unlp.edu.ar}, L. Bronfman$^{2}$, E. Mendoza$^{3}$, M. Merello$^{3}$, R. Finger$^{2}$, 
\newauthor
N. Reyes$^{2,4}$,
C. Herv\'ias-Caimapo$^{5}$,
A. Faure$^{6}$,
C.E. Cappa$^{7}$,
E. M. Arnal$^{7}$,
\newauthor
J.R.D. L\'epine$^{3}$,
I. Kleiner$^{8}$, 
and L-\AA. Nyman$^{9}$\\       
$^{1}$Instituto Argentino de Radioastronom\'{\i}a, CONICET, CCT-La Plata, C.C.5., 1894, Villa Elisa, and CIC, Prov. de Bs. As. Argentina\ \\
$^{2}$Departamento de Astronom{\'{\i}}a, Universidad de Chile, Casilla 36, Santiago de Chile\\
$^{3}$Instituto de Astronomia, Geof\'{\i}sica e Ci\^encias Atmosf\'ericas, Universidad de S\~ao P\~aulo, S\~ao P\~aulo 05508-090, SP, Brazil\\
$^{4}$Departamento de Ingenier\'{\i}a El\'ectrica, Universidad de Chile, Santiago de Chile.\\
$^{5}$Department of Physics, Florida State University, Tallahassee, FL 32306, USA\\  
$^{6}$Universit\'e  Grenoble Alpes, CNRS, IPAG, 38000 Grenoble, France\\
$^{7}$Facultad de Ciencias Astron\'omicas y Geof{\'{\i}}sicas, Universidad Nacional de La Plata, Paseo del Bosque s/n, 1900 La Plata,  Argentina\\
$^{8}$Laboratoire Interuniversitaire des Systemes Atmospheriques (LISA), CNRS, UMR 7583, Universit\'e de Paris-Est Cr\'eteil et Paris Diderot,\\
61, Av G\'en\'eral de Gaulle, 94010 Cr\'eteil Cedex, France\\
$^{9}$Joint ALMA Observatory, JAO Alonso de C\'ordova 31070, Vitacura,  Santiago de Chile, Chile.}

\date{Accepted XXX. Received YYY; in original form ZZZ}

\pubyear{2015}

\begin{document}


\label{firstpage}
\pagerange{\pageref{firstpage}--\pageref{lastpage}}

\maketitle

\begin{abstract}
  
 Using APEX-1 and APEX-2 observations, we have detected and studied the rotational lines of the HC$_3$N molecule  (cyanoacetylene) in the powerful outflow/hot molecular core G331.512-0.103. We identified thirty-one rotational lines  at $J$ levels between 24 and 39; seventeen of them in the ground vibrational state $v$=0 (9 lines  corresponding to the main C isotopologue and  8 lines corresponding to the $^{13}$C isotopologues),    and fourteen  in the  lowest vibrationally excited state $v_7$=1.   Using LTE-based population diagrams for the beam-diluted $v$=0  transitions, we determined $T_{\rm exc}$=85$\pm$4 K and  $N$(HC$_3$N)=(6.9$\pm$0.8)$\times$10$^{14}$ cm$^{-2}$, while  for the beam-diluted $v_7$=1 transitions we obtained $T_{\rm exc}$=89$\pm$10 K and $N$(HC$_3$N)=2$\pm$1$\times$10$^{15}$ cm$^{-2}$.  Non-LTE calculations using H$_2$ collision rates indicate that the HC$_3$N emission is in good agreement with LTE-based results. From the non-LTE method we estimated $T_{\rm kin}$ $\simeq$90~K, $n$(H$_2$)$\simeq$2$\times$10$^7$~cm$^{-3}$ for a central core of 6 arcsec  in size. A vibrational temperature in the range from 130~K to 145~K was also determined, values which are very likely lower limits. Our results suggest that rotational transitions are thermalized, while  IR radiative pumping processes are probably more efficient than collisions in exciting the molecule to the vibrationally excited state $v_7$=1. Abundance ratios derived under LTE conditions for  the $^{13}$C isotopologues suggest that the main formation pathway of HC$_3$N is ${\rm C}_2{\rm H}_2 + {\rm CN} \rightarrow {\rm HC}_3{\rm N} + {\rm H}$.

\end{abstract}




\section{Introduction}


Cyanopolyynes molecules (HC$_{2n+1}$N, $n\geqslant$1) are typical carbon-chain molecules that    seem to be ubiquitous in the interstellar medium (ISM) since they have been detected in a wide range of  environments both  in the Milky Way and in other galaxies. The shortest member of the family, HC$_3$N (cyanoacetylene; H--C$\equiv$C--C$\equiv$N),  is the most abundant cyanopolyyne.  Due to its low rotational constant (1/13 of CO) and large dipole moment (3.7 D, \citealt{dm85}), cyanoacetylene lines can be  detected over a wide range of rotational lines (closely spaced by only 9.1 GHz) at relatively low frequencies, in contrast with, for example, HCN or HCO$^+$, which have high-$J$ lines at very high frequencies that are extremely difficult to detect with ground-based telescopes. 
\begin{figure*}
	\includegraphics[width=15cm]{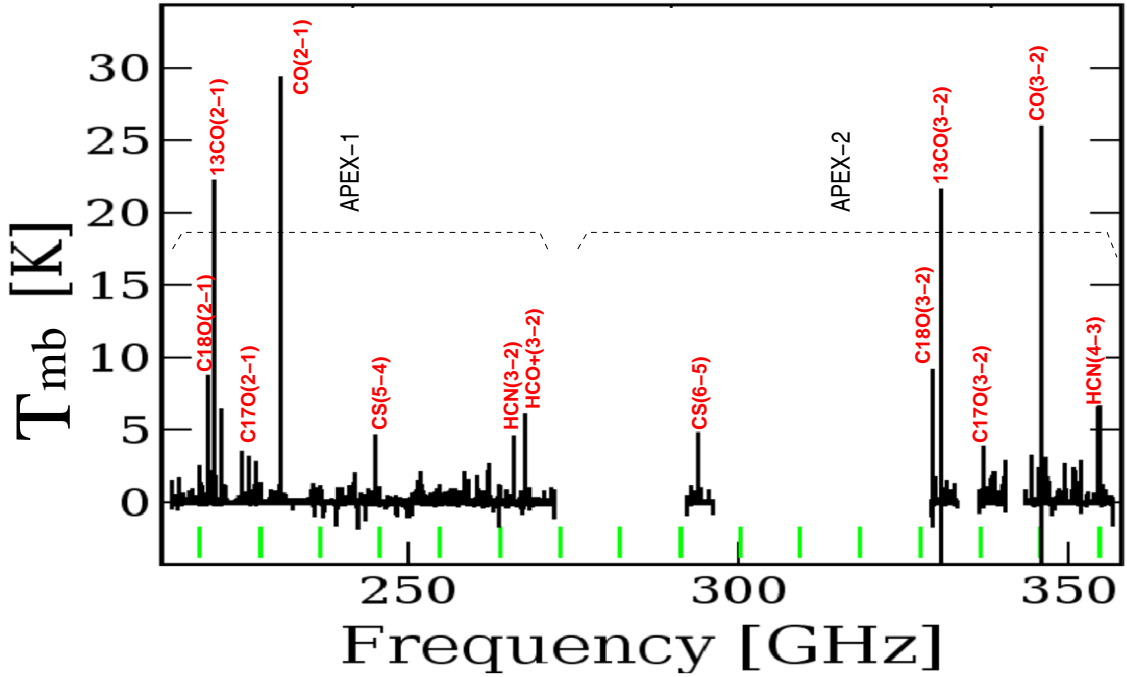} 
 \caption{Total spectrum obtained with APEX-1 and APEX-2 data,  used to identify the HC$_3$N lines. The rest frequencies of the $v$=0 transitions of the  main C isotopologue of HC$_3$N  are indicated by the  green dashes at the bottom of the spectrum. Some of the most prominent and characteristic molecular transitions in the observed bands are indicated in red as a reference. }
 \label{espectrototal}
\end{figure*}
The critical density of HC$_3$N ($n_{\rm H_2}$$\sim$$10^5$-$10^6$ cm$^{-3}$) is comparable to other high density gas tracers, such as  HCN, CS, or HCO$^+$. In addition, HC$_3$N is usually optically thin even in  low-$J$ transitions, mostly due to its  relatively low abundance, which is suitable for studying  dense  molecular  gas (e.g \citealt{ber96}). This makes it easier to conduct multi-transition analysis for HC$_3$N  than for other  dense  molecular  gas  tracers, which can  help  to better  understand the physical and excitation conditions of dense molecular gas in star forming regions. Further,   the HC$_3$N molecule can be excited not only in the ground vibrational state, but also in several vibrationally excited states that can also be used to derive important information about the physical characteristics of the region under study, such as kinetic and vibrational temperatures, density, etc.  \citep{wy99,pint05,jim09,peng17}.

Several  lines of HC$_3$N have been widely detected in a number of hot cores and high-mass star forming regions   \citep{ber96,kim00,dev2000,soll04,bell13,sm13,tani16b,tani18b,tani18c}. Surveys of sources at different evolutionary stages, from high-mass starless cores to protostellar objects, reveal the presence of spectral signatures of both  HC$_3$N and HC$_5$N \citep{tani18}. In the protostellar shock L1157-B1, including observations toward its protostar L1157-mm, lines of HC$_3$N and HC$_5$N were observed showing different physical components of the gas \citep{men2018b}. In the solar-type protostar IRAS 16293-2422, observations of HC$_3$N and HC$_5$N also revealed different gas conditions between the outer and inner regions of the protostellar system \citep{jab2017}. In protoplanetary disks, observations at high spatial resolution of HC$_3$N and CH$_3$CN show abundance ratios of those molecules in agreement with protostellar envelopes and comets  \citep{Bergner2018}.

As part of a project aimed at studying in detail the physical and chemical properties of the massive outflow/hot molecular  core G331.512-0.103 \citep{bro08}, in this paper we present an analysis of the above source  based on  APEX observations of HC$_3$N lines. The excellent spectral and spatial resolution of the  SHeFI APEX-1 and APEX-2 data, along with the obtained spectral coverage, provides a powerful tool to perform a complete analysis on the source.

\section{The molecular outflow/hot molecular core G331.512-0.103 }

G331.512-0.103 (hereafter G331)  is a bright massive young stellar object associated with the central region of the giant molecular cloud (GMC) G331.5-0.1 \citep{bro08,mer13a}, which is located at the tangent point of the Norma spiral arm at  a heliocentric distance of $\sim$7.5 kpc.  The presence of very broad emission wings in CO, CS, and SiO indicates  a very powerful outflow  with a mass of around 55 $M_{\odot}$ and a momentum of $\sim$2.4$\times$10$^3$ \hbox{$M_{\odot}$ km s$^{-1}$}  \citep{bro08}. \citet{mer13b} analyzed ALMA \hbox{SiO(8-7)}, \hbox{H$^{13}$CO$^+$(4-3)}, \hbox{HCO$^+$(4-3)}, and \hbox{CO(3-2)}  observations towards G331  revealing the existence of lobes closely aligned with the line of sight and an expanding bubble blown by stellar winds likely arising from  a hyper compact HII region confined within an angular extent of about  5 arcsec ($\sim$0.18 pc at a distance of 7.5 kpc).

Several transitions of CH$_3$CN and CH$_3$OH  have also been reported in the central core harbouring G331 \citep{men18}.  The results revealed that the emission of CH$_3$CN and CH$_3$OH are related to gas components having kinetic temperatures of T$_{\rm k}$$\sim$141 K and T$_{\rm k}$$\sim$74 K, respectively. In agreement with previous high-resolution interferometric observations, the models show that the emitting region is compact ($\leqslant$ 5.5 arcsec), with a gas volume density \hbox{$n_{\rm H_2} \simeq$1$\times$10$^7$} cm$^{-3}$. 
Recently, ALMA observations have revealed the presence of HC$_3$N $J$=38--37 rotational line \citep{herv19}. The authors identified two group of lines. The first group, showing broad ($\sim$ 20 \kms) velocity wings, includes HC$_3$N, SiO, S$^{18}$O, HCO$^+$, and H$^{13}$CN; very likely they trace the outflow and shocked gas emission. The second group, with narrow \hbox{(6-10 \kms)} emission lines, includes CH$_3$CCH, CH$_3$OH and H$^{13}$CO$^+$ lines. They may be arising from  the quiescent core ambient medium.

 The physical, chemical, and kinematical  properties of G331 make this  object one of the most massive, powerful, and youngest hot cores/outflows known to date.

\section{Observations, databases, and methodology}

The molecular observations presented in this paper were obtained  with the 12-m APEX telescope  \citep{gu06} at Llano de Chajnantor (Chilean Andes) in May 2017. As front end for the observations we used the APEX-1 receiver of the  Swedish Heterodyne Facility Instrument (SHeFI;  \citealt{ri06,vass08}). The back end for the observations was the eXtended bandwidth Fast Fourier Transform Spectrometer2 (XFFTS2)  which consists of two units with 2.5 GHz bandwidth divided into 32768 channels each.  The half power beam width (HPBW) in the APEX-1 frequency band varies between 25 - 30 arcsec.  Calibration was performed using the chopper-wheel technique. The output intensity scale given by the system is $T_{\rm A}^*$, which represents the antenna temperature corrected for atmospheric attenuation.  The observed intensities were converted to main-beam brightness temperature  by $T_{\rm mb}$ = $T_{\rm A}^*$/$\eta_{\rm mb}(\lambda)$, where   $\eta_{\rm mb}(\lambda)$ was obtained using the Ruzes's equation for a mean main beam efficiency of 0.75 and a surface accuracy (rms) of 17 $\mu$m.  The spectra were obtained in the single point mode in the coordinates  \radec = 16$^h$12$^m$10.1$^s$, $-$51\gra28\arcmin38.1\arcsec\ in  sixteen frequency setups to cover the entire APEX-1 band (214 to 272 GHz). Because only up to  4 GHz  total instantaneous bandwidth can be used, frequency setups were spaced by 3.7 GHz in order to avoid noisy marginal detections. The spectra were hanning-smoothed to obtain a final rms noise of \hbox{$\sim$ 0.025 -- 0.05} K and a final spectral resolution of \hbox{$\sim$ 0.4} \kms\  for all the setups.

\begin{figure*}
	\includegraphics[width=12.5cm]{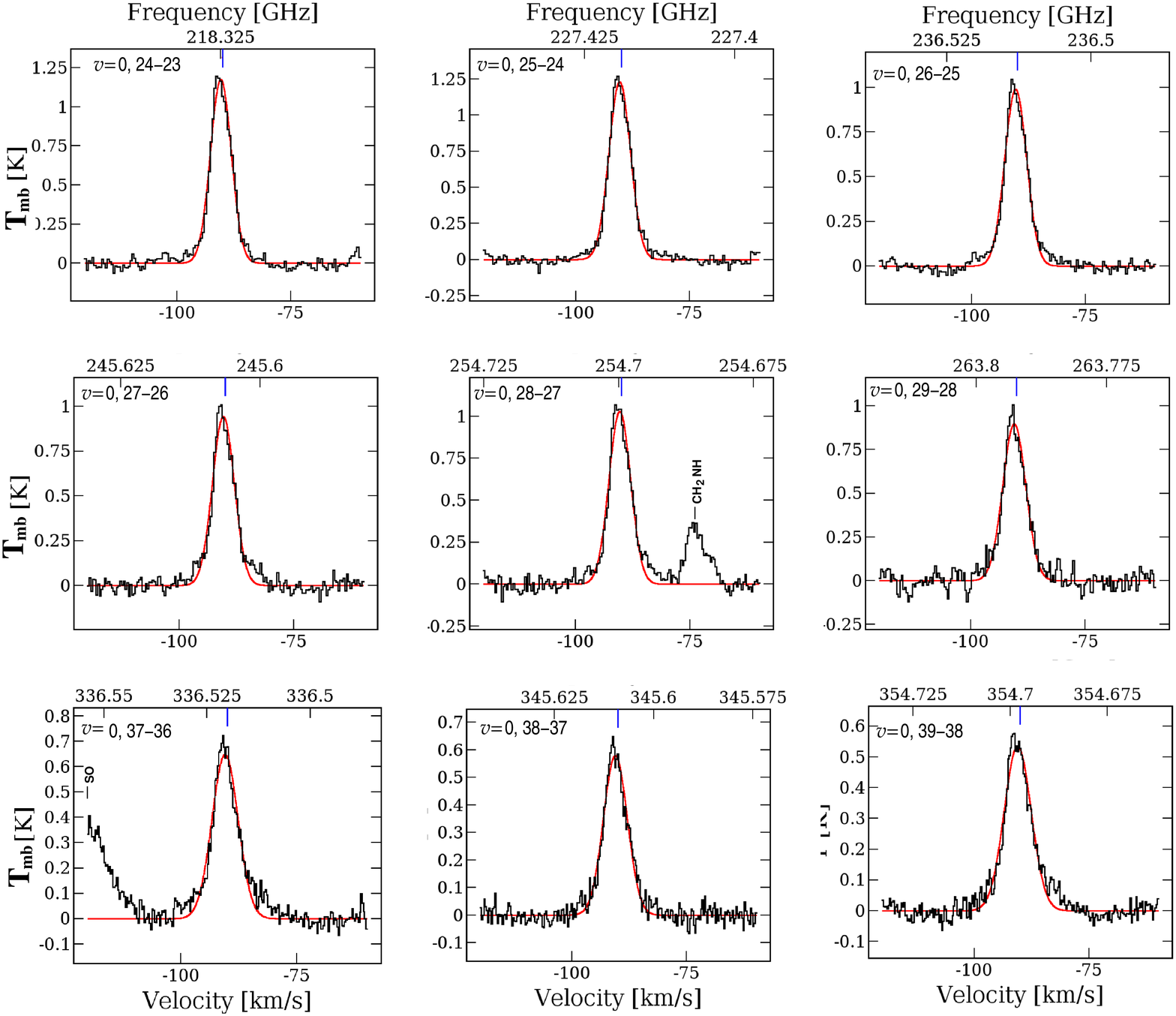}
  \caption{Observed spectra of the detected lines of HC$_3$N  in the ground vibrational state $v$=0. The red curves show the gaussian fit to the lines. The transitions are  indicated in the top left corner of each panel . The blue line at the top of the spectra indicate the rest frequency of each transition.   }
  \label{v=0lines}
\end{figure*}

\begin{figure*} 
	\includegraphics[width=12.5cm]{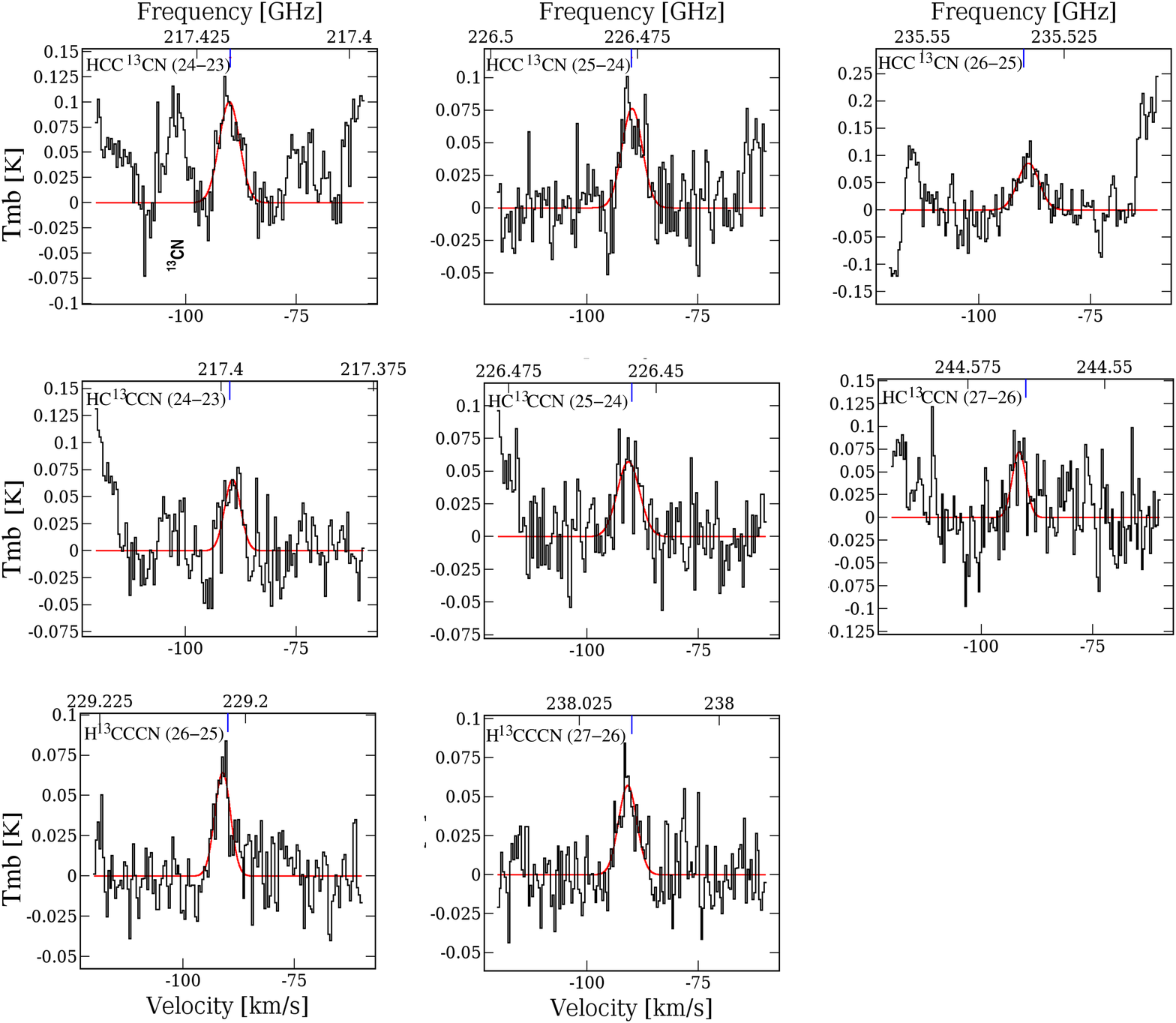}
  \caption{ Observed spectra of the detected  $^{13}$C isotopologues  of HC$_3$N   in the ground vibrational state $v$=0. The red curves show the gaussian fit to the lines. The corresponding isotopologue and its  transition are  indicated in the top left corner of each panel. The blue line at the top of the spectra indicate the rest frequency of each transition.  }
  \label{isotopos}
\end{figure*}

\begin{figure*} 
	\includegraphics[width=16.5cm]{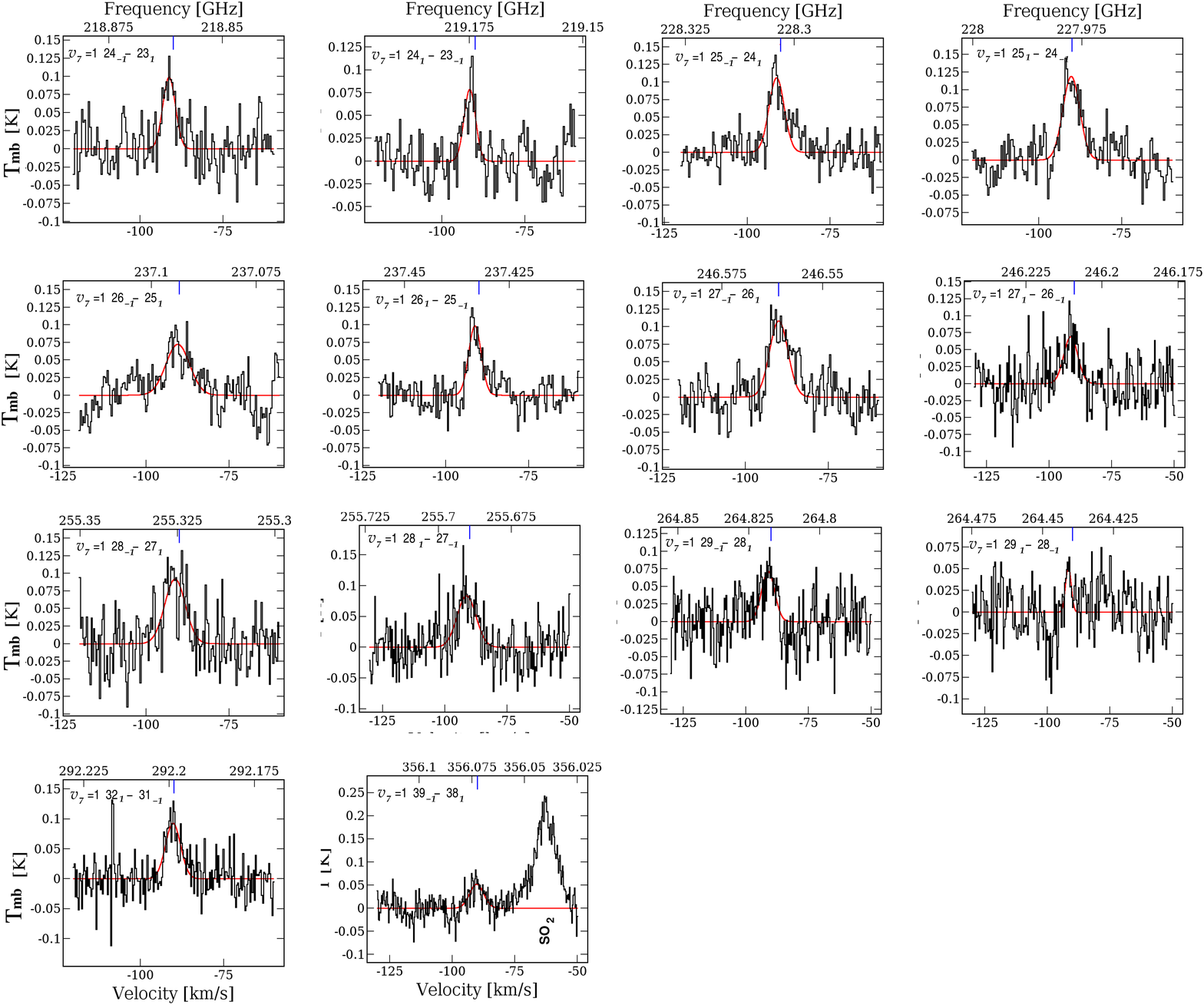} 
  \caption{Observed spectra of the detected HC$_3$N $v_7$=1 lines. The red curves show the gaussian fits to the lines. The transition are  indicated in the top left corner of each panel. The blue line at the top of the spectra indicate the rest frequency of each transition.}
  \label{v=7}
\end{figure*}

\begin{table*}
\caption{HC$_3$N lines detected in G331  and their parameters obtained from Gaussian fits. }
\label{tab:hc3n-tab}
\begin{tabular}{cccccccccc}
\hline  
\hline 
Transition & Frequency & $E_u$ & $A_{ul}$ & $\int T_{\rm mb}\ d{\rm v}$ & V$_{\rm lsr}$& $T_{\rm peak}$  & FWHM & $\tau$   &Remark$^{\S}$ \\
             & (MHz)     &(K) & ($\times$ 10$^{-5}$s$^{-1}$) & (K km s$^{-1}$) &  (km s$^{-1}$) &  (K)  & (km s$^{-1}$) &  & \\
 \hline   
 $v=0$	&		&		&		&			&		&		&	&	\\
\hline
24--23	&	218324.72	&	130.98	&	82.6	&	6.94 $\pm$	0.72  &	-90.35	$\pm$	0.05	& 1.17 $\pm$  0.05	 & 5.39 $\pm$	0.11 & 0.298$^{\ast}$	 &cl\\
25--24	&	227418.91	&	141.89	&	93.5	&	7.49 $\pm$	0.75 &	-90.37	$\pm$	0.05	& 1.23 $\pm$  0.05   & 5.57 $\pm$	0.11 & 0.278$^{\ast}$	 & cl	\\
26--25	&	236512.78	&	153.24	&	105.2	&	6.20 $\pm$  0.62 &	-90.35	$\pm$	0.05	& 0.99 $\pm$  0.04 	& 5.55 $\pm$	0.12 & 0.260$^{\ast}$	 &cl	\\
27--26	&	245606.31	&	165.03	&	117.9	&	6.02 $\pm$	0.61 &	-90.37	$\pm$	0.07	& 0.94 $\pm$  0.06	 & 5.63 $\pm$	0.17 & 0.248$^{\ast}$	 &cl\\
28--27	&	254699.51	&	177.25	&	131.6	&	6.47 $\pm$	0.66 &	-90.37	$\pm$	0.07	&  1.02 $\pm$  0.06  & 5.58 $\pm$	0.15 & 0.235$^{\ast}$  &pbw $^{33}$SO$_2$ \\                            
        &               &           &           &                 &                         &      &              &                & nl CH$_2$NH	\\
29--28	&	263792.31	&	189.91	&	146.3	&	5.85 $\pm$	0.59 &	-90.49	$\pm$	0.07	&	0.89 $\pm$  0.06  & 5.93 $\pm$	0.16 & 0.206$^{\ast}$	 &pbw CH$_3$OH	\\
37--36	&	336520.08	&	306.91	&	304.8	&	4.86 $\pm$	0.50 &	-90.50	$\pm$	0.08	&   0.65 $\pm$  0.05	 & 6.67 $\pm$	0.19 & 0.082$^{\ast}$	 &  nl SO	\\
38--37	&	345609.01	&	323.49	&	330.2	&	3.87 $\pm$	0.39 &	-90.57	$\pm$	0.06	&	0.58 $\pm$  0.04 & 6.16 $\pm$	0.14 & 0.077$^{\ast}$	 &pbw CH$_3$OH	\\
39--38	&	354697.46	&	340.51	&	357.1	&	4.14 $\pm$	0.43 &	-90.54	$\pm$	0.08	&	0.53 $\pm$  0.04   &6.92 $\pm$	0.18 & 0.060$^{\ast}$	 &cl	\\
\hline
$^{13}$C Isotopologues 		&		&		&	&				&		&		&		\\
\hline
H$^{13}$CCCN    &		&		&		&			&		&		&		& \\
\hline
26--25 & 229203.09 & 148.51 & 95.7 & 0.28 $\pm$ 0.04 & -91.1 $\pm$ 0.2	& 0.06 $\pm$  0.02  & 4.1 $\pm$ 0.4 &  0.024$^{\dag}$ & cl    \\
27--26 & 238015.68 & 159.93 & 197.3	& 0.30 $\pm$ 0.04	& -90.9	$\pm$ 0.2 &	0.06 $\pm$  0.02  & 4.8 $\pm$ 0.5 & 0.022$^{\dag}$   & cl	\\
\hline
HC$^{13}$CCN &    &		&		&		&			&		&		&	&	\\
\hline
24--23  & 217398.56	& 130.42 &  81.5 & 0.28 $\pm$ 0.05 & -89.4 $\pm$ 0.4 &  0.07 $\pm$  0.02    & 4.2 $\pm$ 0.8 	&  0.025$^{\dag}$ & cl\\
25--24  & 226454.18	& 141.29 &  92.3 & 0.32 $\pm$ 0.05  & -90.7 $\pm$ 0.4 &  0.06 $\pm$  0.02   & 5.9 $\pm$ 1.0 &  0.024$^{\dag}$  & cl\\
27--26  & 244564.44	& 164.33 &	116.4 &	0.32 $\pm$ 0.06 & -91.5 $\pm$ 0.4 &  0.07 $\pm$  0.03   & 3.6 $\pm$ 0.8 &  0.027$^{\dag}$  &pbw CH$_3$OH	\\
\hline   
HCC$^{13}$CN    &		&		&		&			&		&		&		& &\\
\hline
24--23 & 217419.57 & 130.44	& 81.6	& 0.51 $\pm$ 0.06 & -90.2 $\pm$ 0.2  & 0.10 $\pm$  0.02  & 5.4 $\pm$ 0.4 &	  0.035$^{\dag}$ & pbw CH$_3$OH \\
       &           &        &       &                 &                  &                  &               &                   & nl $^{13}$CN \\
25--24 & 226476.04 & 141.31 & 92.3	& 0.43 $\pm$ 0.06 & -89.9 $\pm$ 0.4  & 0.08 $\pm$  0.03  &5.4 $\pm$ 0.7 &	 0.031$^{\dag}$	& cl\\
26--25 & 235532.20 & 152.61	& 103.9	& 0.50 $\pm$ 0.07 &  -88.9 $\pm$ 0.3 &	 0.09 $\pm$  0.02  & 5.6 $\pm$ 0.6 &	 0.037$^{\dag}$	& cl\\
\hline
$v_7=1$	&		&		&		&			&		&		&		&\\  
\hline
24$_{-1}$--23$_{1}$	&	218860.81	&	452.15	&	82.7  &	0.58 $\pm$	0.07	&	-91.3	$\pm$	 0.3  & 0.10 $\pm$  0.03	&  4.9 $\pm$ 0.5   & 0.015$^{\ast}$	& cl	\\
24$_{1}$--23$_{-1}$	&	219173.75	&	452.34	&	83.0  &	0.34 $\pm$	0.05	&	-91.6	$\pm$	  0.3 & 0.08 $\pm$  0.02	& 4.1	$\pm$ 0.6	& 0.018$^{\ast}$	& cl	\\
25$_{1}$--24$_{-1}$	&	227977.27	&	463.09	&	93.5  &	0.69 $\pm$	0.08	&	-90.2	$\pm$	  0.2 &	0.12 $\pm$  0.02  & 6.0	$\pm$ 0.5	& 0.013$^{\ast}$	& cl	\\
25$_{-1}$--24$_{1}$	&	228303.17	&	463.29	&	93.4  &	0.70 $\pm$	0.09	&	-90.8	$\pm$	0.3	 &  0.10 $\pm$  0.02  & 6.8	$\pm$ 0.6	& 0.011$^{\ast}$	& cl	\\
26$_{-1}$--25$_{1}$	&	237093.38	&	474.47	&	105.2 &	0.53 $\pm$  0.07	&	-90.5	$\pm$	0.4	 &	0.07 $\pm$  0.02   & 8.4	$\pm$ 1.2	& 0.009$^{\ast}$	& cl	\\
26$_{1}$--25$_{-1}$	&	237432.26	&	474.69	&	105.7 &	0.57 $\pm$	0.07	&	-91.0	$\pm$	0.2	 &	0.10 $\pm$  0.03  & 5.3 $\pm$ 0.5	& 0.014$^{\ast}$	& cl	\\
27$_{1}$--26$_{-1}$	&	246209.14	&	486.29	&	118.0 &	0.45 $\pm$	0.08	&	-91.4	$\pm$	0.5	 &	0.07 $\pm$  0.03  & 9.0	$\pm$ 1.9	& 0.011$^{\ast}$	& cl	\\
27$_{-1}$--26$_{1}$	&	246560.95	&	486.52	&	118.5 &	0.78 $\pm$	0.10	&	-89.3	$\pm$	0.4	 &	0.10 $\pm$  0.03 & 8.1	$\pm$ 0.9	& 0.009$^{\ast}$	& cl	\\
28$_{-1}$--27$_{1}$	&	255324.55	&	498.54	&	131.7 &	0.58 $\pm$	0.10	&	-91.5	$\pm$	0.5	 &	0.09 $\pm$  0.03 & 6.7	$\pm$ 1.1	& 0.011$^{\ast}$	& cl	\\
28$_{1}$--27$_{-1}$	&	255689.29	&	498.79	&	132.2 &	0.72 $\pm$	0.11	&	-91.3	$\pm$	0.6	 & 0.08 $\pm$   0.02	 & 8.9	$\pm$ 1.6	& 0.008$^{\ast}$	& cl	\\
29$_{1}$--28$_{-1}$	&	264439.58	&	511.23	&	146.4 &	0.20 $\pm$	0.04	&	-91.6	$\pm$	0.3	 & 0.05 $\pm$   0.03	 & 2.6	$\pm$ 0.7	& 0.029$^{\ast}$	&	cl \\
29$_{-1}$--28$_{1}$	&	264817.24	&	511.50	&	147.1 &	0.40 $\pm$	0.07	&	-91.1	$\pm$	0.5	 & 0.07 $\pm$   0.02	  & 6.2	$\pm$ 1.1	& 0.012$^{\ast}$	&	cl  \\
32$_{1}$--31$_{-1}$	&	292198.64	&	552.26	&	197.9 &	0.50 $\pm$	0.08	&	-90.2	$\pm$	0.3	 & 0.09 $\pm$   0.02	& 4.9	$\pm$ 0.6	& 0.014$^{\ast}$	& cl	\\
39$_{-1}$--38$_{1}$	&	356072.44	&	662.68	&	359.2 &	0.31 $\pm$	0.06	&	-90.3	$\pm$	0.4	 & 0.05 $\pm$   0.02	& 6.6	$\pm$ 1.0	& 0.008$^{\ast}$	& nlo SO$_2$	\\
\hline

\hline
\end{tabular}

$^{\ast}$Obtained iteratively from the opacity correction (Sect 4.2.1)
$^{\dag}$Obtained from Eq.\ref{tau} (Sect. 4.2.3)\\
\end{table*}
The spectra were reduced using the CLASS90 program of the IRAM GILDAS software package\footnote{http://www.iram.fr/IRAMFR/GILDAS}. The analysis of the spectra and the line identification was performed using the CASSIS software\footnote{http://cassis.irap.omp.eu} conjointly  with the CDMS\footnote{https://www.astro.uni-koeln.de/cdms} \citep{mul05}, JPL\footnote{https://spec.jpl.nasa.gov/} \citep{pick98}, and NIST\footnote{https://physics.nist.gov/cgi-bin/micro/table5/start.pl} \citep{lov09} spectroscopic databases.

The APEX-1 observations used in this work were combined with previous observations carried out using the APEX-2 instrument (HPBW = 17 - 23 arcsec)  (see Fig.~\ref{espectrototal}). We adopted calibration uncertainties of around 20 percent \citep{dum10}.




\section{Results and analysis}

\subsection{ Line identification}

The APEX-1 and APEX-2 data used in this work cover the total frequency range between 214 GHz  and 356.5 GHz,  with the APEX-1 frequency range between 214 and 272 fully covered and some  gaps in the APEX-2 band between 272 and 292 GHz, 296 and 329 GHz, 333 and 336.4, and 340.4 and 343.3 GHz (see Fig.~\ref{espectrototal}). Nevertheless, the covered frequency range allows  the   detection of a considerable number of HC$_3$N lines, having an upper level energy  ($E_u$) of the main isotopologue between $\sim$131 K and  $\sim$341 K. The detection of these lines is necessary  to perform a high accuracy  determination of some  physical parameters in the region (e.g. column density, kinetic temperature, opacity, etc.).

Since the central region of G331 shows a very rich hot molecular core chemistry, a forest of molecular transitions leads to a confusing spectra (see Fig.~\ref{espectrototal}), especially  for the presence of broad lines \citep{men18,herv19}.  Then, we performed the line identification adopting  the following standard criteria: 
\begin{enumerate}
\item The observed frequencies of the lines should agree with the rest frequencies for the systemic velocity of the source  (V$_{\rm lsr}$ $\sim$ --90 \kms)\footnote{In this work radial velocities are always referred to the Local Standard of Rest (lsr)}. \\

\item The  peak temperature of the lines should be  at least  about three times the  rms noise. A couple of  exceptions were included where the signal-to-noise ratio is  locally a bit lower  than in the rest of the spectrum. \\

\item All predicted lines based on an LTE spectrum at a well-defined rotational temperature  should be present in the observed spectrum approximately at their predicted relative intensities.
\end{enumerate}

Since a  considerable number of predicted contaminant molecular transitions, known as weeds\footnote{Term usually adopted to refer to species with numerous spectral lines in the mm/sub-mm regime}, can be detected at several frequencies in the mm and sub-mm range, we also inspected for possible contaminant emission. Such inspection was carried out using spectral catalogs and the Weeds extension of the CLASS radio astronomy software.\footnote{\url{https://www.iram.fr/IRAMFR/GILDAS/doc/html/weeds-html/weeds.html}} Thus, we remarked three status in the line identification (see last column of Table~\ref{tab:hc3n-tab}):  a) lines possibly blended with known interstellar species (pbw): transitions whose FWHM values derived from the Gaussian fits coincide with predicted frequencies of interstellar species detected in the ISM  (in all the cases the possible contaminat  species were detected in the ISM but at different transitions, therefore significant contamination is unlikely).    b) neighbor lines (nl): criterion that was used to identify the spectral lines which appear within a spectral window equivalent to $V_{\rm lsr}$~-90 $\pm$ 30~km~s$^{-1}$; c) clean lines (cl): lines whose spectral profiles and FHWH values are not affected by other spectral features, such as blended or neighbor lines.


\subsubsection{$v$=0 lines} 

Using the criteria mentioned above, we  confirmed the detection of  9 rotational lines of the main C isotopologue of HC$_3$N in the ground vibrational state ($v$=0). Six of them  were   identified   in the APEX-1 band, while three  lines were identified in the APEX-2 band (see Fig.~\ref{v=0lines}).  Identified lines  are listed in Table~\ref{tab:hc3n-tab} along with some of their spectroscopic parameters like frequency,  energy of the upper level of the transition ($E_u$), and Einstein coefficient of the transition ($A_{ul}$). For these lines we  performed a Gaussian fit and  we obtained their integrated areas ($\int T_{\rm mb}\ d{\rm v}$), linewidths (FWHM), peak temperatures ($T_{\rm peak}$), and velocities (V$_{\rm lsr}$). These parameters are also indicated in Table~\ref{tab:hc3n-tab}. As can be seen from the table, the velocity dispersion    of the peak velocities of the lines around the systemic velocity  V$_{\rm lsr}$ = $-$90 \kms\  is quite low \hbox{($\leqslant$ 0.57 \kms)}, and the mean systemic velocity of the lines is $\bar{\rm V}_{\rm lsr}$=--90.43$\pm$0.08 \kms.  In Table~\ref{tab:hc3n-tab} we also indicate the optical depth of each line ($\tau$) that will be further derived in Section~4.2.1.


   It is also worth mentioning that  the frequency range analysed in this work covers several transitions of larger cyanopolines (thirty-one for the molecule  HC$_5$N and over forty for the molecule HC$_7$N) and still,  no lines corresponding to these species were detected in our dataset.

\subsubsection{Isotopologues}
 
 We also searched for lines of the $^{13}$C-, D-, and $^{15}$N-bearing isotopologues of HC$_3$N that are predicted to be present in the observed frequency intervals. We identified $^{13}$C isotopologue lines with maximum $J$ numbers between 24 and 27, all in the ground vibrational state $v$=0. They are shown in  Fig.~\ref{isotopos}.   The transition (25--24) of H$^{13}$CCCN was not detected since it overlaps with the strong $^{13}$CO \hbox{(2--1)} line. The same occurs with the (26--25) transition of HC$^{13}$CCN, which seems to overlap with another strong ($T_{\rm mb}$ $\sim$ 1.3 K) unidentified line. Thus, a complete triad H$^{13}$CCCN--HC$^{13}$CCN--HCC$^{13}$CN could not be detected for a particular $J$ transition. As a marginal detection, we included  the HC$^{13}$CCN (27--26) line, which is better identified over an extra hanning-smoothed spectra (not shown here). 
 
 The velocity dispersion of the derived peak velocity of the lines with respect to the systemic velocity   \hbox{($\leqslant$ 1.5 \kms)}  is larger than that of the main isotopologue, probably due to a lower accuracy of the fit. The mean systemic velocity of the lines is $\bar{\rm V}_{\rm lsr}$=--90.3$\pm$0.8  \kms.     No transitions for the  DC$_3$N and HC$_3^{15}$N isotopologues  were detected in our  dataset.

\subsubsection{$v_7$=1 lines}

   The molecule of HC$_3$N  has seven modes of vibration, three bending modes (doubly degenerate) and four stretching modes, which are usually identified from $v_1$  to $v_7$  (for a complete revision, see \citealt{wy99}). Then, besides rotational lines in the ground vibrational state ($v$=0), we also searched for rotational transitions of HC$_3$N in vibrationally excited states. As a result, we found transitions  in  the  lowest excited  state, $v_7$=1 (corresponding to the bending vibration mode of the \hbox{C--C$\equiv$C} group). The detected lines  are shown in Fig.~\ref{v=7}. The interaction between  the bending angular momentum of the vibrational $v_7$=1 state and the rotational angular momentum of the molecule leads to a splitting in two rotational levels for each $J$$\rightarrow$$J-1$ transition (indicated in Fig.~\ref{v=7} as -1 and 1).  As for the case of the HC$^{13}$CCN (27--26) line (see Sect.~4.1.2), we included the marginal detection of the transition \hbox{29$_1$--28$_{-1}$}. It should be noted that the transition \hbox{39$_1$--38$_{-1}$} was not detected since it is blended with an intense ($\sim$ 0.25 K) and wide ($\sim$ 8 \kms) unidentified line at $\sim$ 355.569 GHz.   The spectroscopic and Gaussian fit parameters are listed in Table~\ref{tab:hc3n-tab}. In particular, most of them have the same transitional levels of HC$_3$N $v$=0  with maximum $J$ numbers between 24 and 39.  A velocity dispersion of $\leq$1.6 \kms\ for the derived peak velocities of the lines is observed around the systemic velocity of the source, which has a mean value of  $\bar{\rm V}_{\rm lsr}$=--90.9$\pm$0.7  \kms. For consistency, we performed the Gaussian fit to the lines using the velocity interval defined for the \hbox{25$_1$--24$_{-1}$} transition, which is the most conspicuous one.

\subsection{Line analysis}

\subsubsection{LTE analysis} 
\label{lte-analysis}

In order to estimate excitation temperatures, $T_{\rm exc}$, and column densities, $N$(HC$_3$N), we constructed population diagrams from the identified transitions of the molecule. This term refers to a plot of the column density per statistical weight of a number of transition energy levels, as a function of their energy above the ground state \citep{gol99}. In local thermodynamical equilibrium (LTE), and assuming that the emission is optically thin, this corresponds to the Boltzmann distribution and can be represented as
\begin{equation}
\ \ \ \ \ {\rm ln}\left(\frac{N_u}{g_u}\right) = {\rm ln}\left(\frac{N_{\rm tot}}{Q(T_{\rm exc})}\right)-\frac{E_u}{kT_{\rm exc}},
\label{diagrot}
\end{equation}
 where $N_u/g_u$ is the column density per statistical weight for the upper level $u$,   $E_u$ is the energy of the  upper level, $Q(T_{\rm exc})$ is the partition function, and $N_{\rm tot}$ is the total column density. Since $N_u$ can be estimated from the integrated intensity of the line as
\begin{equation}
\ \ \ \ \ N_u =   \frac{8 \pi k \nu^2}{h c^3 A_{ul}}    \times \int{T_{\rm mb} \  d{\rm v}}  
\label{nu} 
\end{equation}
then, Eq.~\ref{diagrot} represents a Boltzmann distribution whose values ln$(N_u/g_u)$ vs. $E_u$ can be obtained from the lines and can be fitted using a straight line, whose slope is defined by the term 1/$T_{\rm exc}$. 

The population diagrams were constructed using a  beam dilution factor  ${\rm ln} \left(\frac{\Delta\Omega_s}{\Delta\Omega_a + \Delta\Omega_s}\right)$  where $\Delta\Omega_a$ and $\Delta\Omega_s$ are the solid angle of the antenna beam and the solid angle of the source, respectively. The beam dilution factor was applied on the left side of Eq.~\ref{diagrot}. A source size of $\sim$6 arcsec was adopted (see Sect.~4.2.2).

 \begin{figure}  
	\includegraphics[width=8.7cm]{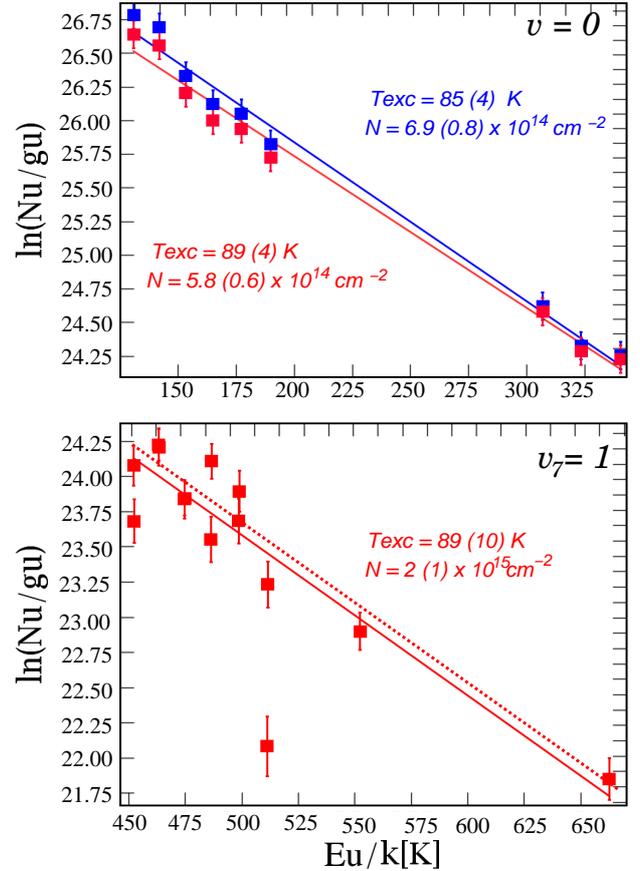} 
  \caption{{\it Upper panel}: Population diagrams for the HC$_3$N $v$=0 lines, showing the least squares fits without optical depth correction (red color) and with optical depth correction (blue color). The corresponding errors in density and temperature are indicated in parenthesis. {\it Lower panel}: Population diagrams for the HC$_3$N $v_7$=1 lines. The upper dotted line represents the least square fit without the transition line 29$_{1}$-28$_{-1}$ (see transitions with $E_u$=511 K). }
  \label{DR} 
\end{figure}

In the upper panel of Fig.~\ref{DR} we show the obtained population diagrams for the HC$_3$N $v$=0 lines. From the excellent fit obtained from the data ($\chi^2_{red}$=1.28) we derived $N$(HC$_3$N)=(5.8$\pm$0.6)$\times$10$^{14}$ cm$^{-2}$  and $T_{\rm exc}$=89$\pm$4~K.  The plotted uncertainties in Fig.~\ref{DR} are:
\begin{equation}
\ \ \ \ \ \Delta \left({\rm ln}\ \frac{N_u}{g_u}\right) = \frac{\Delta W}{W},
\label{error}
\end{equation}
were $W$ is the integrated area of the line and $\Delta W$ is its uncertainty,  estimated as\footnote{http://cassis.irap.omp.eu/docs/RadiativeTransfer.pdf}
\begin{equation}
\ \ \ \ \ {\Delta W} = \sqrt{\ (cal/100 \times W)^2 + \left(rms  \sqrt{\ 2 \times {\rm FWHM} \times \Delta{\rm v} } \right)^2}
\label{w}
\end{equation}
being $cal$ the calibration error (in percentage), $rms$ the noise around the selected line (in K), FWHM the line full width half maximum (in km s$^{-1}$), and $\Delta$v the bin size (in km s$^{-1}$)


The population diagrams obtained before were constructed under assumption of optically thin emission \hbox{($\tau<<$1)}.  In order to test the reliability of this approximation, we have introduced the optical depth correction factor $C_{\tau}$   (photon escape probability) in Eq.~\ref{nu} as
\begin{equation}
\ \ \ \ \   N_u =  \frac{8 \pi k \nu^2}{h c^3 A_{ul}} \times C_{\tau} \times \int{T_{\rm mb} \  d{\rm v}}
\label{nuc}
\end{equation}
where $C_{\tau}$=$\tau/(1 - e^{-\tau})$ \citep{gol99}, being $\tau$ the optical depth of the line. When the line is optically thin, the factor  $C_{\tau}$ is then equal to unity. Thus,  Eq.~\ref{diagrot} becomes
\begin{equation}
\ \ \ \ \ {\rm ln}\left(\frac{N_u}{g_u}\right) = {\rm ln}\left(\frac{N_{\rm tot}}{Q(T_{\rm exc})}\right)-{\rm ln}\ C_{\tau}-\frac{E_u}{kT_{\rm exc}},
\label{diagrotC}
\end{equation}
and the ordinate of the population diagram fit is rescaled to slightly larger values. The correction is performed with CASSIS, which execute iteratively the calculations of $C_{\tau}$ until values of  temperature and column density converge. From the fit  ($\chi^2_{red}$=1.33) we derived a column density $N$(HC$_3$N)=(6.9$\pm$0.8)$\times$10$^{14}$ cm$^{-2}$  and a excitation temperature $T_{\rm exc}$=85$\pm$4 K. Then, the percentage differences of $N$(HC$_3$N) and $T_{\rm exc}$ with respect to the values obtained  without optical depth correction are about       19 percent higher and 5 percent lower, respectively. For the sake of comparison, the corrected population diagram is shown  in Fig.~\ref{DR} together with the original diagram without optical depth correction. The optical depth of each line, estimated with this method, is listed in Table~\ref{tab:hc3n-tab}. As can be noticed  from Table~\ref{tab:hc3n-tab} and Fig.~\ref{DR}, differences are moderately significant ($\tau$ $\gtrsim$ 0.2)  only for low $J$-values  ($J<28$). This trend is in line with  the variation of the optical depth as a function of $J$ presented  by \citet{gol99}  for  the molecule of HC$_3$N with $h$B/$kT$=0.003, for which a maximum relative optical depth is obtained for $J$ $\approx$ 20, and monotonically decreasing for $J$$\lesssim$20 and $J$$\gtrsim$20 (see fig.~1 of that work).    
    
As can be noticed from Fig.~\ref{v=0lines} some lines show Lorentzian instead of Gaussian profiles, which is a very usual feature in emission lines affected by outflows.   After  a Lorentzian fit of the beam-diluted lines, we derived an opacity-corrected column density $N$(HC$_3$N)=(7.8$\pm$0.5)$\times$10$^{14}$ cm$^{-2}$  and an excitation temperature $T_{\rm exc}$=90$\pm$3 K, respectively, which means a difference of $\sim$13 percent higher in density and  $\sim$7 percent also higher  in temperature, compared with values obtained by  Gaussian fitting.


In the lower panel of Fig.~\ref{DR} we also show the population diagrams obtained for the $v_7$=1  lines.  From the fit we derived a column density and temperature of  $N$(HC$_3$N)=(2.2$ \pm$2.0)$\times$10$^{15}$ cm$^{-2}$  and $T_{\rm exc}$=87$\pm$14 K, respectively. The low accuracy of the fit ($\chi^2_{red}$=6.07)   is due, not only to the significantly lower signal-to-noise ratio of the detections, but mostly to the  transition line 29$_{1}$-28$_{-1}$ ($E_u = 511.23$ K) which lies outside the 3 rms scatter limit of the plot. Removing this line from the plot, we obtained a more accurate fit ($\chi^2_{red}$=2.8); the  column density and  temperature obtained then are    $N$(HC$_3$N)=(2$\pm$1)$\times$10$^{15}$ cm$^{-2}$  and $T_{\rm exc}$=89$\pm$10 K, respectively. As expected, these values are consistent with those obtained for the $v$=0 lines. Optical depths obtained iteratively for the $v_7$=1 lines are $<$0.03 (see Table~\ref{tab:hc3n-tab}), which  amounts to  a  difference of only 0.6 percent and 3.5 percent in temperature and  column density, respectively, with respect to those obtained without  opacity correction (not shown in Fig.~\ref{DR}). The excitation temperatures derived before for the rotational transitions in the $v$=0 and $v_7$=1 states are in good agreement with those derived from the CH$_3$OH emission by \citet{men18} and from CH$_3$CCH emission by \citet{herv19} for   the ambient  gas in G331.  We note, however, that HC$_3$N column densities derived above for G331 are almost one order of magnitude larger than some found in typical Galactic hot molecular cores (e.g. \citealt{tani18,tani18c}).

Besides  transitions of different $J$ levels in the same vibrational state, the level population of different $v$ states for a constant rotational $J$ level can also be analyzed   to derive the so-called "vibrational temperature" ($T_{\rm vib}$) which determines the relative population of  vibrationally excited  levels. A rough estimate of $T_{\rm vib}$ can be obtained from  a comparision between vibrationally excited lines  with ground-state lines (e.g. \citealt{gold82}). Here, assuming optically thin emission, we use  
\begin{equation}
\ \ \ \ \ \ \frac{W_{(v_{7}=1)}}{W_{(v=0)}} = {\rm exp}\ \left( \frac{- \Delta E_{7 0}}{k T_{\rm vib}} \right)  
\label{tvib}
\end{equation}
where $W_{(v_{7}=1)}$ and $W_{(v=0)}$ stand for  the integrated intensities of a $J$$\rightarrow$$J$-1 transition for the vibrational states $v_{7}$=1 and $v$=0, respectively, and  $\Delta E_{7 0}$ is the energy difference between the $v$=0 and $v_7$=1 states. Thus, we estimated $T_{\rm vib}$ for  constant $J$ levels in the range between 24 to 27 and we obtained vibrational temperatures  in the range  130 K to 145 K. As expected, obtained values of $T_{\rm vib}$ are higher than those  previously obtained, for rotational transitions within the same vibrational state, from the population diagrams (a similar trend was also reported by \citealt{gold82} in the Orion Molecular Cloud  using $v_7$=1 and $v$=0 lines of HC$_3$N) but lower than those reported in the literature  for Galactic hot cores and  star forming regions (e.g. \citealt{wy99,jim09,peng17}).  It is worth mentioning that the above  values of $T_{\rm vib}$ should be considered as lower limits since the $v$=0 state is expected to be  dominated by more extended and cooler emission. Thus,  the ratio $W_{(v_{7}=1)}$/$W_{(v=0)}$ is rather a measure of the relative population in the $v_7$=1 vibrational state instead of a measure of $T_{\rm vib}$ \citep{wy99}. Observations of more vibrational modes, either bending or stretching, are needed in order to determine $T_{\rm vib}$ more accurately.   

   \subsubsection{Non-LTE analysis using collision rates}
 \label{non-lte-analysis}

  All the estimates obtained so far rely on the assumption of LTE regime. By means of the excitation diagrams displayed in Fig.~\ref{DR}, we obtained single excitation temperatures from the whole set of observed rotational lines of HC$_3$N, both in  ground and vibrationally excited states. The Boltzmann distribution of the excitation diagrams and the quality of the straight lines (with slope 1/$T_{\rm exc}$, Eq.~\ref{diagrotC}) seem to validate the LTE condition for rotational transitions of  HC$_3$N. As a consequence, the kinetic temperatures would be expected to be similar to the excitation temperatures if actually all the  levels  were  thermalized.

For a better understanding of the excitation mechanisms and optical depth effects, as to determine the gas kinetic temperature and density, it would also be useful to carry out an analysis of the HC$_3$N emission assuming non-LTE conditions. To this end, we use statistical equilibrium calculations to perform non-LTE models of the HC$_3$N ($\nu$=0) lines. The statistical equilibrium calculations were carried out using the RADEX code\footnote{\url{https://home.strw.leidenuniv.nl/~moldata/radex.html}} \citep{van2007}. Then, by means of non-local radiation effects we tested how far is the HC$_3$N emission from the LTE hypothesis considering collisional excitations induced by molecular hydrogen, whose rate coefficients are presented and discussed in \citet{fau2016}. The RADEX calculations were computed simultaneously applying the Markov Chain Monte Carlo method (MCMC) \citep{Foreman2013},\footnote{\url{http://dfm.io/emcee/current/}} which is hereafter referred to as RADEX/MCMC. In general, the RADEX/MCMC method computes the best solution by means of numerical chains in a n-dimensional parameter space that can be delimited by different free parameters, such as the kinetic temperature ($T_{\rm kin}$), column density ($N$), linewidths (FWHM), source size, and volume density ($n$(H$_2$)). Taking into account the availability of collision rate coefficients, the last parameter is set accordingly to the H$_2$-HC$_3$N collision system (e.g. \citealt{fau2016}). Through stochastic processes, the RADEX/MCMC finds the best solution up to achieve the minimum value of the $\chi^2$ function. Since the method operates using the spectral lines as main input, the quality of the models depends on the spectral survey (e.g. number and strength of the lines), the free parameters and their ranges of integration. 

\begin{figure*} 
	\includegraphics[width=15.5cm]{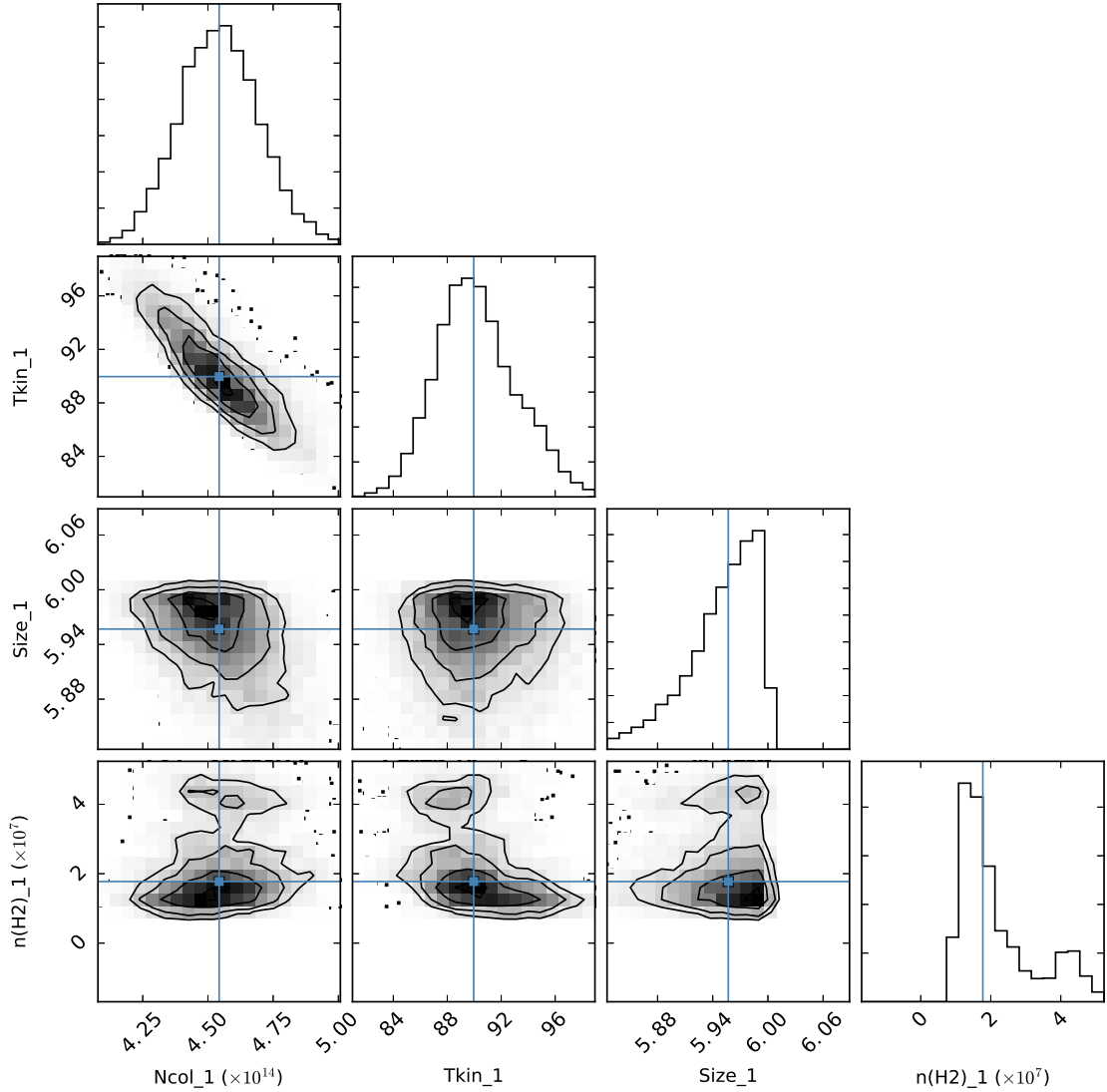}
  \caption{Corner diagram with the highest probability distribution of the column density (Ncol, cm$^{-2}$), kinetic temperature (Tkin, K), source size (Size, arcsec) and H$_2$ density (n(H2),  cm$^{-3}$) derived from the best RADEX/MCMC model ($\chi^2_{red}$=4.41) of HC$_3$N. Collision excitation coefficients between HC$_3$N and H$_2$ were used to perform the calculations considering the HC$_3$N lines from $J$=24--23 to $J$=29--28. The blue solid lines indicate the median values.}
  \label{corner}
\end{figure*}

 The RADEX/MCMC models of HC$_3$N were computed considering the first six rotational lines listed in Table~\ref{tab:hc3n-tab}, from $J$=24--23 to $J$=29--28. As free parameters, we introduced and calculated the distributions of $T_{\rm kin}$,  $N$(HC$_3$N), $n({\rm H}_2)$, and source size. In order to encompass the HC$_3$N emission,  the statistical equilibrium calculations  were carried out assuming an emitting region $\leq$  6 arcsec in size,  a value consistent with the HC$_3$N \hbox{$J$=38--37} emission map exhibited in \citet{herv19}.  Since the quality of the computation is a monotonically increasing function of the numbers of steps or iterations, we included in our routines large samples for draw numbers, of the order of 10$^4$. 




As a corner diagram, the RADEX/MCMC results of HC$_3$N are shown in Fig.~\ref{corner}. In that plot, the histograms of each free parameter merge into  bi-dimensional diagrams exhibiting the highest probability distribution. Distributions of the HC$_3$N column density, H$_2$ density, kinetic temperature and source size are exhibited.   Summarizing, the physical conditions derived from the best RADEX/MCMC computation ($\chi^2_{red}$=4.41), were obtained to be $N$(HC$_3$N)=4.5$\pm$0.1$\times$10$^{14}$~cm$^{-2}$ and $T_{\rm kin}$=90$\pm$3~K for a central core with size and H$_2$ volume density of $\lesssim$ 6 arcsec and $\sim$ 2$\times$10$^7$~cm$^{-3}$, respectively.


 In comparison with the LTE analysis of HC$_3$N, the results obtained assuming non-LTE conditions are  relatively in good agreement  with those obtained by  the excitation diagram for  the $v$=0 state (see Sect.~4.2.1). The percentage differences of column density and temperature with respect to the values obtained in LTE conditions for the $v$=0 lines with optical depth correction are about  6 percent higher  and 53 percent lower,  respectively, while the difference with respect to the values without optical depth correction is only about 1 percent higher and 28 percent lower, respectively.      


\subsubsection{Isotopic analysis}

Since a scarce number of transitions from isotopic species were detected,  and some of them  having a  poor signal-to-noise ratio,  a population diagram does not provide reliable solutions for the temperature and column density. Then, we estimated the column density of the three detected  $^{13}$C isotopologues from a single line of each species assuming LTE conditions  and using the following formulae: 
\begin{equation}
\ \ \ \tau\ =\ -{\rm ln}\left[1 - \frac{T_{\rm mb}}{f\ [J(T_{\rm exc})-J(T_{\rm bg})]} \right]
\label{tau}
\end{equation}
being
\begin{equation}
\ \ \ J(T) = \frac{h \nu}{k} \left[e^{\left(\frac{h \nu}{k T}\right)} -1 \right]^{-1},   
\end{equation}
and
\begin{equation}
N = \tau\  \frac{3h\Delta v}{8\pi^3}\ \sqrt{\frac{\pi}{4\ {\rm ln}2}}\ \frac{Q}{\mu^2 (J_l + 1)} \ \ e^{\left(\frac{E_l}{k T_{\rm exc}}\right)} \left[ 1 - e^{\left(-\frac{h \nu}{k T_{\rm exc}}  \right)}    \right]^{-1} 
\label{N} 
\end{equation}
\citep{gol99}, where $T_{\rm mb}$ stands for the main beam temperature of the line, $f$ for the beam filling factor (estimated to be $\sim$ 0.035\footnote{Estimated considering a source size of $\sim$ 6 arcsec (see Sect.~4.2.2) and an average beam size of $\sim$ 27 arcsec for the APEX-1 band}),  $T_{\rm exc}$ is the excitation temperature (assumed to be 85 K; see Sect. 4.2.1), and $T_{\rm bg}$ is the background temperature (2.7 K). For this equation, we estimate an error of about \hbox{15--25} percent arising mostly from the uncertainties  in $T_{\rm mb}$ (see Table~\ref{tab:hc3n-tab}). In Eq.~\ref{N} $\Delta v$ is the line width FWHM (in cm s$^{-1}$), $Q$ is the partition function estimated as $Q$ = $kT_{\rm exc}$/$h$B (B is the rotational constant of HC$_3$N; 4549.059 MHz), $\mu$ is the permanent electric dipole of HC$_3$N (assumed to be 3.73172$\times$10$^{-18}$ esu cm; \citealt{dm85}), $E_l$ (in erg) is the energy of the lower rotational level, and $J_l$ is the lower rotational quantum number. Obtained column density for each isotopologue is given  in Table \ref{tab:isotopos}, while optical depths obtained with Eq. \ref{tau} are listed  in Table~\ref{tab:hc3n-tab}.  The errors estimated for the column density of the $^{13}$C isotopologues were derived considering the uncertainties in $\tau$ (see above) and $\Delta v$ (see Table \ref{tab:hc3n-tab}).    For the calculations, we used the most conspicuous line detected for each isotopic species. Then, for HCC$^{13}$CN and HC$^{13}$CCN we used the (24--23) transition, while for H$^{13}$CCCN we used the (26--25) transition.


\begin{table}
\caption{ Physical properties derived for the HC$_3$N and its $^{13}$C isotopologues. } 
\label{tab:isotopos}
\begin{tabular}{lcccc}   
\hline
 Isotopologue &   $N$    &    $X$   &   {$^{12}$C/$^{13}$C}    \\
              & [$\times$ 10$^{13}$ cm$^{-2}$] &  [$\times$ 10$^{-11}$] &     \\
\hline 
HC$_3$N      &  69  $\pm$ 8$^{\dag}$      &  79 -- 62   &    -  \\
HCC$^{13}$CN &   6.7 $\pm$ 0.9    &   7.8 -- 5.9  &   $\sim$ 10 \\
HC$^{13}$CCN &   3.8 $\pm$ 0.8    &   4.7 -- 3.1  &   $\sim$ 18 \\
H$^{13}$CCCN &   3.7 $\pm$ 0.6    &   4.4 -- 3.2  &   $\sim$ 19 \\
$^{13}$C HC$_3$N  &   3.3 $\pm$ 0.2$^{\ddag}$   & 3.6 -- 3.2 & $\sim$ 21 \\
\hline
\end{tabular} 
$^{\dag}$Obtained by population diagram (see Sect.~4.2.1)  \\
$^{\ddag}$Obtained by the LTE/MCMC method \\
\end{table}


The relative abundances of the isotopic species with respect to H$_2$ were derived using  the H$_2$ column density obtained previously from the emission of H$^{13}$CO$^+$ (e.g. \citealt{mer13a,men18}). ALMA observations of the \hbox{H$^{13}$CO$^+$(4-3)} emission have shown the existence of an internal cavity surrounded by a shocked shell, around the driving source, confined in $\lesssim$ 5 arcsec \citep{mer13a} for which an averaged column density $N$(H$^{13}$CO$^{+}$)$\approx$3.2$\pm$0.2$\times$10$^{13}$ cm$^{-2}$ was derived (result obtained by  \citealt{herv19}, but not included in that work). Then, we estimated the H$_2$ column density using the abundance ratio H$^{13}$CO$^{+}$/H$_2$=3.3$\times$10$^{-11}$ of Orion KL \citep{bl87} and we obtained $N$(H$_2$)=9.7$\times$10$^{23}$ cm$^{-2}$, almost one order of magnitude higher than that derived in typical high-mass star forming regions (e.g. \citealt{tani18b}).  The abundances of HC$_3$N, HCC$^{13}$CN, HC$^{13}$CCN, and H$^{13}$CCN were then obtained dividing their column densities by $N$(H$_2$). In Table~\ref{tab:isotopos} we summarize the range of obtained abundances,  indicated as $X$, which were derived considering the uncertainties of  column densities estimated using Eq.~\ref{N}. It is worth pointing out that $N$(H$_2$) can be otherwise estimated from the surface density derived by \citet{mer13a} from the 1.2 mm continuum emission of G331 (referred to as MM3 in that work). Adopting a mean molecular weight per Hydrogen  molecule $\mu$=2.29 (corrected for Helium), a column density $N$(H$_2$)=2.7$\times$10$^{23}$ cm$^{-2}$ can be obtained, which would increase the abundances reported in Table~\ref{tab:isotopos} by a factor of $\sim$3.5 

Since HC$_3$N and its $^{13}$C isotopologue lines are optically thin (see Col.~9 of Table~\ref{tab:hc3n-tab}) we can obtain reliable measurements of the $^{12}$C/$^{13}$C isotopic ratio in G331 from the column densities calculated before. For the estimations we have considered  column densities of HC$_3$N and its isotopologues obtained  in LTE conditions. The obtained $^{12}$C/$^{13}$C isotopic ratios are shown in the last column of  Table~\ref{tab:isotopos}.  

   \begin{figure} 
  \centering
	\includegraphics[width=9cm]{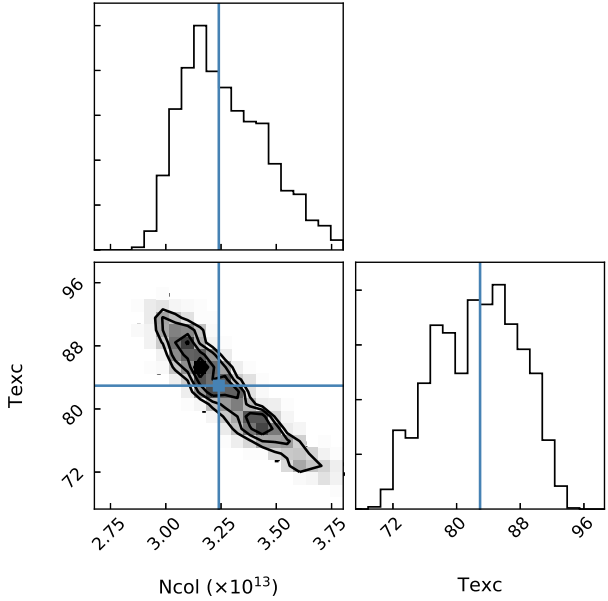}
  \caption{Corner diagram with the highest probability distributions of the column density N$_{\rm col}$ (cm$^{-2}$) against the excitation temperature T$_{\rm exc}$ (K) derived from the LTE/MCMC models of the $^{13}$C HC$_3$N isotopologues. The best model ($\chi^2_{red}$=5.88) yielded $T_{\rm exc}$=82$\pm$5~K and $N$(H$^{13}$CCCN)=$N$(HC$^{13}$CCN)=$N$(HCC$^{13}$CN)=(3.3$\pm$0.2)$\times$10$^{13}$~cm$^{-2}$  for a source size of 6 arcsec.  The blue solid lines indicate the median values.}
  \label{13C}
\end{figure}

 In addition to the estimates obtained above, we also  carried out LTE models of the $^{13}$C isotopologues of HC$_3$N  applying the MCMC method, whose methodology will hereafter referred to as LTE/MCMC. From the lines of the isotopologues listed in Table~\ref{tab:hc3n-tab}, we obtained LTE solutions under the main assumption that the $^{13}$C HC$_3$N isotopologues and HC$_3$N have the same origin, (i.e. they are confined in a region of up to 6 arcsec and with $T_{\rm exc}$ between $\sim$~80~K and 90~K). The LTE/MCMC models were computed including as free parameters only the column density and excitation temperature. As an upper limit, based on the premise that HC$_3$N must be more abundant than its isotopologues, the $^{13}$C column densities were integrated with the condition that  $N<N$(HC$_3$N) $\thickapprox$ 1$\times$10$^{15}$~cm$^{-2}$.  A single LTE/MCMC computation was carried out for all the $^{13}$C isotopologues  to obtain a single excitation solution for all of them. Such computation included all the observed lines of the $^{13}$C isotopologues (Table~\ref{tab:hc3n-tab}). We found that the best model ($\chi^2_{red}$=5.88) to reproduce the observed lines of H$^{13}$CCCN, HC$^{13}$CCN and HCC$^{13}$CN  is the one that assumes $N$=(3.3$\pm$0.2)$\times$10$^{13}$~cm$^{-2}$ and $T_{\rm exc}$=82$\pm$5~K for a source size of 6 arcsec. The LTE/MCMC results are shown in Fig~\ref{13C} as a corner diagram, where the highest probability distribution of $T_{\rm exc}$ against $N$ is exhibited. A second approach was made by adopting  $T_{\rm}$=85~K (see Sect.~4.2.1) and computing only the column density as a free parameter. The best model ($\chi^2_{red}$=5.87) yielded  $N$=(3.18$\pm$0.04)$\times$10$^{13}$ cm$^{-2}$, which is lower than the column density reported above only by a very small percentage, namely $\sim$ 4 percent.

\section{Discussion}

\subsection{Excitation for the v=0 and v7=1 states}

In Sects. 4.2.1 and 4.2.2, we used LTE and non-LTE approaches to derive the density and temperature in G331. Obtained values for $T_{\rm k}$ and $N({\rm HC_3N}$) from both methods are in good agreement for the $v$=0 lines which suggests that detected transitions are almost thermalized ($T_{\rm exc}$=$T_{\rm kin}$). However, it is expected that low-energy transitions may sample the emission from both hot and cold gas components, while highly excited transitions are expected to be rather originated from hot gas environments. Thus, high rotational transitions  may also be excited by radiative process, which  become  more efficient as the angular momentum of the upper state increases (high-$J$ levels).  

The critical density of a rotational transition $J$$\rightarrow$$J$-1 can be estimated  as
\begin{equation}
\ \ \ \ \ \ n_{\rm c}^{J,J-1} \approx \frac{A_{(J,J-1)}}{\gamma{(J,J-1)}}
 \label{ncrit}
\end{equation}
being $\gamma_{(J,J-1)}$ the collisional coefficient for the $J$$\rightarrow$$J$-1 transition, and $A_{(J,J-1)}$ its Einstein's transition probability. If $n$(H$_2$) $<<$ $n_{\rm c}^{J,J-1}$ the excitation temperature is mostly determined by the nearby radiation field instead of the gas kinetic temperature, which implies that the $J$ level is not thermalized ($T_{\rm exc}$ < $T_{\rm kin}$). Using  the  exclusive collisional rate coefficients discussed in \citet{fau2016}, we derived critical densities for the $J$=38$\rightarrow$37 transition for temperatures between 80 K and 100 K and we obtained values between 1.6$\times$10$^7$ cm$^{-3}$ and 1.3$\times$10$^{7}$ cm$^{-3}$. These values are a bit lower than that estimated in Sect. 4.2.2 ($\sim$ 2$\times$10$^7$ cm$^{-3}$), which very likely confirms that the gas density in G331 is high enough for thermalizing the molecule, even at the higher $J$ levels detected in our observations. 

The detection of vibrationally  excited lines ($v_7$=1) suggests that the excitation of the molecule might be strongly influenced, besides collisions, by radiative processes. As previously done in the above paragraph, we must compare the ambient density derived in Sect. 4.2.2, with the critical density for the vibrational bending mode $v_7$=1. The latter was estimated by \citet{wy99}, who derived $n_{\rm c}$ for the $v_7$=1-0 transition of HC$_3$N using the far-IR absorption measurements of \citet{uye74} to estimate the Einstein's transition probability ($A_{u,l}$), and using semiempirical formula for vibrational relaxation times as a function of temperature, reduced mass, and energy vibration \citep{gold82}, to estimate the collisional coefficient. Then, the authors derived  $n_{\rm c}$ = 4 $\times$ 10$^8$ cm$^{-3}$ for $T = 300$ K (a suitable temperature for a hot core). This value is almost one order of magnitude higher than the ambient  density derived for G331 in this work. 
This implies that IR radiative excitation processes (likely originated by the thermal emission of dust grains)  are  probably more efficient than collisions in exciting the vibrationally upper state $v_7$=1 of the molecule.

 In case of IR radiative excitation and assuming that the gas and dust are thermally coupled (a good approximation considering a gas density of 2$\times$10$^7$ cm$^{-3}$; see Sect.~4.2.2) the temperature $T_{\rm vib}$ determined from the HC$_3$N emission is a good estimate of the dust temperature. Then, we can estimate the IR bolometric  luminosity of the hot core by using the Stefan-Boltzmann law (e.g. \citealt{dev2000}) 


\begin{equation}
\ \ \ \ \ \frac{L_{\rm hc}}{L_{\odot}} = 1.8625\times10^{-5}  \left(\frac{r}{10^{16}\  {\rm cm}}\right)^2\  T_{\rm dust}^4
 \label{lir}
\end{equation}
 were $r$ is the radius of the hot core, and $T_{\rm dust}$ is the dust temperature. Adopting an upper limit emission size of 6 arcsec ($\sim$0.22 pc at a distance of 7.5 kpc) and a temperature $T_{\rm dust}$=130 K (see Sect.~4.2.1) we obtained for G331 an IR bolometric luminosity $L \approx$ 6$\times$10$^6$ $L_{\odot}$. This value    would correspond to a ZAMS star type earlier  than O4 \citep{pana73}.  In order to test this, we  estimated the number of ionizing Lyman continuum photons needed to sustain the ionization in G331 using 
\begin{equation}
\ \ \ N_{\rm Lyc} = 4.5 \times 10^{48}\   T_e^{-0.45}\  S_{\rm (4800\ MHz)}\ d^2
\end{equation}
 \citep{simp90} were $d$ is the distance in kpc, $T_e$ is the electron temperature (assumed as 10$^4$ K), and  $S_{\rm (4800\ MHz)}$ is the radio continuum flux at 4800 MHz in Jy (adopted as 0.095 Jy;  \citealt{mer13b}). Keeping in mind that more than 50 percent  of the UV photons can be absorbed by the ISM  \citep{inoue01}, the total amount of ionizing photons needed to sustain the current level of ionization in G331 is \hbox{$N_{\rm Lyc}$ $\approx$ 1 $\times$ 10$^{48}$ s$^{-1}$}. Adopting fluxes extracted from \citet{pana73}, we estimate the spectral type of the ionizing source to be about O9/O8.5 (ZAMS). Then, the spectral type of the internal stellar source derived by this method  is different than that derived from the IR bolometric luminosity. It should be noted, however, that the latter  magnitude is highly dependant on the size of the hot core (see Eq.~\ref{lir}), which for the case of G331 is highly  uncertain due to the presence of a powerful outflow.


\subsection{12C/13C  ratio}

\subsubsection{Isotopic fractionation and  astrochemical implications}   

As can be seen from Fig.~\ref{isotopos}, the $J$=24$-$23 and $J$=25$-$24 transition lines of the HCC$^{13}$CN  isotopologue are  brighter than those of the HC$^{13}$CCN. This  trend  can be confirmed from their integrated intensities ($\int{T_{\rm mb}\  d{\rm v}}$;  see Table\ref{tab:hc3n-tab}).  The derived ratio HCC$^{13}$CN/HC$^{13}$CCN of the integrated intensities for the $J$=24$-$23 line is  1.8 $\pm$ 0.5, while for the $J$=25$-$24 line is 1.3 $\pm$ 0.4.   On the other hand, the $J$=27$-$26 lines for the H$^{13}$CCCN and HC$^{13}$CCN isotopologues have about the same peak temperature (within the errors), a tendency that can also be confirmed by the ratio H$^{13}$CCCN/HC$^{13}$CCN  integrated intensities, derived to be 0.9 $\pm$ 0.3. Then, observed integrated intensities of the HCC$^{13}$CN lines are stronger by a factor between 1.8 to 1.3 than those of HC$^{13}$CCN lines, while there is no significant difference in intensities between   H$^{13}$CCCN and HC$^{13}$CCN  lines. This kind of anomaly, known as  {\it isotopic fractionation}, has been reported and studied, not only in HC$_3$N but also in HC$_5$N and other carbon-chain molecules,  in several hot and cold sources  (e.g. \citealt{taka98,sak07,sak13,li16,ara16,tani16a,tani16b,tani17b}). 
  
  The relative abundances ratios of the $^{13}$C isotopologues can be also calculated using column densities estimated from the most conspicuous transitions assuming LTE conditions  (see Table~\ref{tab:isotopos}). Thus, the abundance ratios were obtained to be   [HCC$^{13}$CN]:[HC$^{13}$CCN]:[H$^{13}$CCCN] = 1.8($\pm$0.5):1.0:1.0($\pm$0.4). These results seem to confirm  that the abundance of H$^{13}$CCCN and HC$^{13}$CCN are comparable with each other, while HCC$^{13}$CN is more abundant than the others. Similar abundance ratios  of HC$_3$N were reported  in several star-forming and starless cores in the Galaxy  (L1527: \citealt{tani16b,ara16}; TMC-1: \citealt{taka98}; G28.28-0.36: \citealt{tani16b}; Serpens South 1A: \citealt{li16})  
  

 Since the isotope exchange reactions  seem unlikely in the case of HC$_3$N    \citep{taka98}, it has been proposed that the differences in abundances in $^{13}$C isotopologues are  rather a consequence of the formation pathway of the molecule. In the case of HC$_3$N  it has been proposed that the neutral-neutral reaction 
 \begin{equation}
 \label{reactionhc3n}
 \quad {\rm C}_2{\rm H}_2 + {\rm CN} \rightarrow {\rm HC}_3{\rm N} + {\rm H}
 \end{equation}
 is the most important chemical pathway to produce HC$_3$N in the cold cloud TMC-1 \citep{herbst90,taka98}.  Further, \citet{chap09}  demonstrated that this formation pathway also occurs relatively easy at hot core temperatures (100 -- 200 K).  Thus, the chemical source of the $^{13}$C atom can be supplied from C$_2$H$_2$ and/or CN favoring  a given position where the $^{13}$C atom could be located into the molecule. A reaction between predominant $^{13}$C-enriched CN  with  less  enriched $^{13}$C C$_2$H$_2$ in the environs of G331 could then explain the observed  abundance ratios   \hbox{[HCC$^{13}$CN]:[HC$^{13}$CCN]:[H$^{13}$CCCN]} = \hbox{a:b:c}, with a$>$b$\approx$c (e.g.\citealt{tani16b,ara16}). It is worth noting that a transition of the molecule $^{13}$CN at 217.42856 GHz was identified in our dataset, next to the HCC$^{13}$CN (24--23) transition line (see Fig.~\ref{isotopos}). 
 
 We accept, however, that the HC$_3$N abundance ratios reported before have large errors.  Then, it is important to contrast these results with a multitransition study of rotational lines  having larger signal-to-noise ratios (presumably at lower $J$ levels). A complete  analysis of APEX data at lower frequencies will be carried out soon. This study would allow us, not only to estimate the abundance ratios of HC$_3$N more accurately, but also to  detect some transitions of larger cyanopolyynes (e.g. HC$_5$N) not detected in our dataset  (see below).


\subsubsection{Average $^{12}$C/$^{13}$C  ratio}

As shown in Sect.~4.2.3, the average $^{12}$C/$^{13}$C ratio was estimated from LTE/MCMC calculations by computing all the isotopologue lines in a single run, which yielded to a ratio of $\sim$21 (see last line in Table~\ref{tab:isotopos}). Alternatively, the average  $^{12}$C/$^{13}$C ratio of HC$_3$N in G331 can be also  estimated from the column density of each isotopic species as
\begin{equation}   
\label{rav}
^{12}{\rm C}/^{13}{\rm C}_{\rm (HC_3N)} = \frac{3 \times  N_{\rm HC_3N}}{N_{\rm H^{13}CCCN} + N_{\rm HC^{13}CCN} + N_{\rm HCC^{13}CN} }
\end{equation}
 Using Eq.~\ref{rav} and column densities of each isotopic species indicated in  Table~\ref{tab:isotopos}, the average isotopic ratio  was obtained to be $\sim$15.   This result is in relatively good agreement with the $^{12}$C/$^{13}$C ratios derived from observations of CH$_3$OH and CH$_3$CN in G331 \citep{men18}.
 \begin{figure} 
  \centering
	\includegraphics[width=8.5cm]{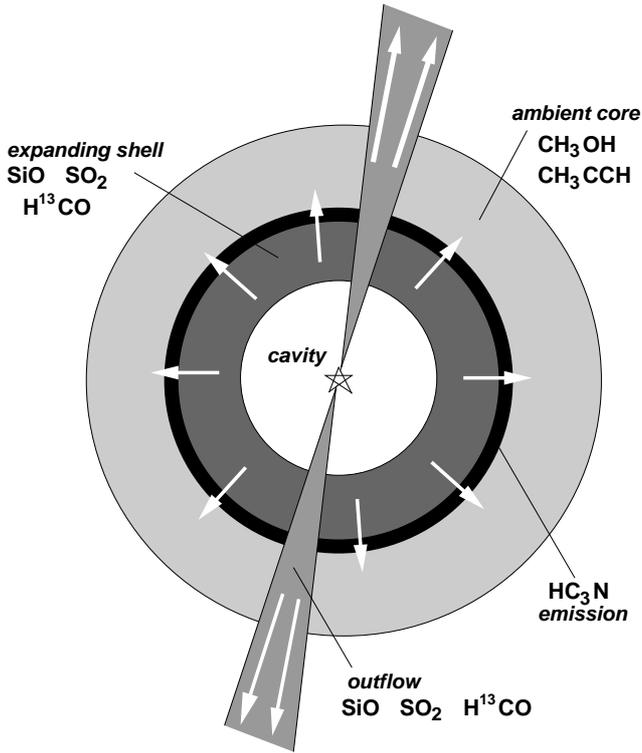}
  \caption{Sketch of the simple model proposed to explain the emission of HC$_3$N and other  molecules detected in G331 (based on the models of \citealt{mer13a} and \citealt{herv19}).}
  \label{modelo}
\end{figure}
 
 In order to compare the  $^{12}$C/$^{13}$C abundance ratio derived for  G331 with that expected from its environs  (in this case, the giant molecular cloud G331.5-0.1) we used the empirical formula   $^{12}$C/$^{13}$C=6.21$\times D_{\rm GC}$+18.71 (derived by \citealt{mil05} for Galactic GMCs) where $D_{\rm GC}$ is the distance of G331.5-0.1 to the Galactic center (derived  using simple trigonometry). Thus,  the expected $^{12}$C/$^{13}$C abundance ratio for the the molecular cloud G331.5-0.1 is $\sim$ 43,  more than twice of that derived for  G331.

\subsection{The emission of  HC$_3$N and larger cyanopolyynes} 

As previously mentioned in Sect.~2, high spatial resolution ALMA observations towards G331 revealed the emission of  SiO\hbox{(8-7)}, H$^{13}$CO$^+$\hbox{(4-3)}, HCO$^+$\hbox{(4-3)}, and CO\hbox{(3-2)} towards the center of the objects \citep{mer13a}. The authors interpreted the emission like a high-velocity jet lying almost along the line of sight with an expanding shocked shell/bubble surrounding the driving source, which is observed in projection as a ring of $\sim$ 5 arcsec in size, at the systemic velocity \hbox{($\sim$ --90 \kms)}. The  line $J$=38--37 of  HC$_3$N was also  reported towards G331 with ALMA observations \citep{herv19}. The authors identified this line as a part of a group of lines (HC$_3$N, SiO, S$^{18}$O, HCO$^+$, and H$^{13}$CN) with broad velocity wings tracing the shocked gas and the outflow, for which a temperature of 160--200 K was derived. The second group (CH$_3$OH, CH$_3$CCH, and H$^{13}$CO$^+$) have  narrow emission lines and trace a colder ($\sim$70~K) core ambient medium. 


 In Sect.~4.2, from HC$_3$N lines, we derived temperatures between $\sim$80~K and 90~K, which  are much closer to the temperature of the (quiescent) ambient gas derived by \citet{herv19} than the one  of the (expanding) molecular shocked gas. This might indicate that the molecular gas traced by  HC$_3$N  may be  physically associated with molecular gas of the ambient core,  which has not been fully disturbed by the plowing power of the  expanding shell. This could  in turn explain the narrower profile of the HC$_3$N line in the group of lines with broad wings (see fig.~1 from \citealt{herv19}).
 
 Thus, we can speculate that the emission of HC$_3$N could be originated in a middle region, between the quiescent ambient core and the expanding molecular gas. Then, the molecular gas traced by the HC$_3$N emission might be being disturbed by the expansion of the shell in its inner region, while its undisturbed outer layer preserves much of the physical properties and systemic velocity of the ambient core (see Fig.~\ref{modelo}).  This is a fairly plausible scenario considering that  G331 has shown to be one of the youngest hot cores/outflows known to date. A  more detailed analysis of high resolution data is necessary to shed some light on these issues.

   
  Regarding  to larger cyanopolyynes, as previously mentioned in Sect.~4, we did not report  evidences for HC$_5$N, HC$_7$N, nor HC$_9$N (e.g. \citealt{Sakai2008,jab2017,Zhang2017,Cordiner2017,men2018b,Seo2019}). The absence of HC$_5$N in the current dataset could be associated with several factors, such as (1) insufficient sensitivity  in the frequencies of the APEX-1 and APEX-2 data, (2) the excitation regime of the molecular environment, or (3)  lower abundances of the molecule. For instance, the second  HC$_5$N ($v$=0) line that should appear in our observations corresponds to the  $J$=82--81  transition with $E_u$=434.8~K near the transition of HC$_3$N ($v$=0) appeared at the level $J$=24--23 with $E_u$=130.98~K (see Table\ref{tab:hc3n-tab}). However, in order to conclude on the detection of HC$_5$N in G331 subsequent studies are needed at lower frequencies with instruments such as APEX SEPIA-B5, or with high resolution ALMA Bands 3, 4 and 5. The frequency interval of the mentioned  ALMA bands, from 84--211 GHz, seems suitable to look for HC$_5$N emission considering the physical and chemical implications of HC$_3$N in G331. For instance, \citet{Agundez2017} using ALMA data discussed the spatial and chemical correlations between HC$_3$N and HC$_5$N in sources such as IRC+10216.

\section{Summary} 

As part of a project aimed at studying the chemical and physical properties prevailing in the massive outflow/hot molecular core G331.512-0.103, we used APEX-1 and APEX-2 observation to detect and analyze the emission of the simplest cyanopolyyne molecule HC$_3$N, a typical hot core tracer. The main results can be  summarized as follows:

\begin{enumerate}
   
   \item We detected thirty-one rotational lines of HC$_3$N above  $\sim$ 3 rms  (with the exception of two marginal cases) at $J$ levels between 24 and 39. Seventeen of them correspond to the ground vibrational state $v$=0 (inlcluding 8 lines of $^{13}$C isotopologues)    and fourteen  correspond to the lowest vibrationally excited state $v_7$=1.\\
    
    \item From multitransition LTE analyses for the $v$=0 and  $v_7$=1 beam-diluted lines  we estimated  excitation temperatures between $\sim$ 85 and 89 K, and HC$_3$N column densities between $\sim$6.9$\times$10$^{14}$ cm$^{-2}$ and    2$\times$10$^{15}$ cm$^{-2}$. From a non-LTE MCMC/RADEX analysis of the $v$=0 lines, we determined that the emission of HC$_3$N is originated in a region of up to 6 arcsec  in size, having  an ambient H$_2$ density of $\sim$2$\times$10$^7$ cm$^{-3}$. The results obtained  from both, LTE and non-LTE approaches, suggest that detected rotational transitions are almost thermalized ($T_{\rm exc} \thickapprox T_{\rm kin}$), even at high $J$ levels. \\

    \item  Using $v$=0 and $v_7$=1 lines we determined  vibrational temperatures between 130 and 145 K, which are very likely lower limits.  The critical density of the $v_7$=1 states indicates that  IR radiative pumping processes are likely  to be more efficient than collisions in exciting the vibrationally upper states of the molecule.\\
    
    \item Using a single-transition LTE analysis,  we estimated the column densities of the $^{13}$C isotopologues of HC$_3$N. Derived values are (6.7$\pm$0.9) $\times$ 10$^{13}$ cm$^{-2}$, (3.8$\pm$0.8) $\times $10$^{13}$ cm$^{-2}$, and (3.7$\pm$0.6) $\times$ 10$^{13}$  cm$^{-2}$, for HCC$^{13}$CN, HC$^{13}$CCN, and H$^{13}$CCCN, respectively. Then, derived densities (and integrated intensities of different lines) yields an abundance ratio of about     [HCC$^{13}$CN]:[HC$^{13}$CCN]:[H$^{13}$CCCN] = 1.8($\pm$0.5):1.0:1.0($\pm$0.4).  By means of a multi-transition analysis using the LTE/MCMC method, an average solution with $N$ = (3.3$\pm$0.2) $\times$ 10$^{13}$~cm$^{-2}$ and $T_{\rm exc}$ = 82$\pm$5~K was also determined for the $^{13}$C HC$_3$N isotopologues. The $^{12}$C/$^{13}$C ratios derived from the observations are in agreement with those previously derived in G331 from CH$_3$OH and CH$_3$CN.
    The evidences for $^{13}$C fractionation reported in this work are in line with the widely accepted chemical formation pathway C$_2$H$_2$+CN$\rightarrow$HC$_3$N+H. \\ 

     \item Based on previous ALMA observations and recent results, a very simple physical model was proposed to attempt to shed some light on the emission of HC$_3$N and other molecules previously detected with ALMA.   
    
\end{enumerate}

\section*{Acknowledgements}
We very much acknowledge the anonymous referee for his/her helpful comments and suggestions that led to the improvement  of this paper. This project was partially financed by CONICET of Argentina under projects PIP 00356 and  from UNLP, 11/G139. L.B. acknowledges support from CONICYT project Basal AFB-170002.




\bibliographystyle{mnras}
\bibliography{bibliografia.bib} 









\bsp	
\label{lastpage}
\end{document}